\RequirePackage{fix-cm}
\documentclass[twocolumn]{svjour3}          
\smartqed  
\usepackage[pdftex]{graphicx}
\usepackage{natbib}
\setcitestyle{aysep={}}
\usepackage{color}
\usepackage{amsmath}
\usepackage{textcomp,gensymb}
\usepackage{comment}

\usepackage{ulem}
\usepackage{upgreek}
\usepackage[none]{hyphenat}
\usepackage{gensymb}

\usepackage{ulem}

 \journalname{}
\usepackage[]{hyperref}
\usepackage{xcolor}
\hypersetup{
    colorlinks,
    linkcolor = blue,
    citecolor = blue,
    urlcolor = blue
}

\newcounter{aq}
\definecolor{darkgreen}{RGB}{0, 128, 0}

{}

\begin{document}
\title{Measurement of flow birefringence induced by the shear components along the optical axis using a parallel-plate-type rheometer}
\subtitle{}

\author{William Kai Alexander Worby$^{1}$ $\boldsymbol{\cdot}$ Kento Nakamine$^{2}$ $\boldsymbol{\cdot}$ Yuto Yokoyama$^{2}$ $\boldsymbol{\cdot}$ \\Masakazu Muto$^{3*}$ $\boldsymbol{\cdot}$ Yoshiyuki Tagawa$^{2**}$}

\authorrunning{William Kai Alexander Worby, Kento Nakamine, Yuto Yokoyama, Masakazu Muto, Yoshiyuki Tagawa}

\institute{
1: Department of Industrial Technology and Innovation, Tokyo University of Agriculture and Technology, Koganei, Tokyo, 184-8588, Japan\\
2: Department of Mechanical Systems Engineering, Tokyo University of Agriculture and Technology, Koganei, Tokyo, 184-8588, Japan\\
3: Department of Electrical and Mechanical Engineering, Nagoya Institute of Technology, Nagoya, Aichi, 466-8555, Japan\\
          $^{*\ }$
        \email{muto.masakazu@nitech.ac.jp}\\
        $^{**}$
        \email{tagawayo@cc.tuat.ac.jp}\\
}
\date{}
\vspace{-10mm}

\maketitle
\sloppy
\begin{abstract}
The present study investigated the flow birefringence induced by shear components along a camera's optical axis, which has been neglected in conventional theories of photoelastic measurements.
Measurements were conducted for a wide range of shear rates from a direction perpendicular to the shear using a high-speed polarization camera and a parallel-plate-type rheometer.
The measurement results obtained from a fluid with low viscoelasticity, specifically a dilute suspension of cellulose nanocrystals, showed that the birefringence increases monotonically as the stress components along the camera's optical axis increase.
It was also found that the birefringence showed a power law with respect to the shear rate.
This letter reports a key fact required for polarization measurements of shear rate (shear stress) in three-dimensional flows.
\end{abstract}



\vspace{-8mm}
\section{Introduction}
\vspace{-4mm}
Measurement of three-dimensional stress fields in fluids is of interest in various disciplines, such as flow engineering, polymer chemistry, and biomechanics.
In particular, flow birefringence can be used for non-invasive stress measurement.
The relationship between the flow birefringence $\Delta_n$ and fluid strain rate $\dot e_{ij}$ is described by \citep{Doyle1982}:
\begin{equation}
\begin{split}
\Delta_n\rm {cos}2&\phi=\alpha_1({\dot e_{xx}}-{ \dot e_{yy}})\\&+\alpha_2[(\dot e_{xx}+\dot e_{yy})(\dot e_{xx}-\dot e_{yy})+\dot e_{zy}^2-\dot e_{xz}^2],
\end{split}
\label{eq:sol-A}
\end{equation}
\begin{equation}
\Delta_n\rm{sin}2\phi=2\it{\alpha_{\rm1}\dot e_{xy}+\alpha_{\rm 2}[\rm{2}\it(\dot e_{xx}+\dot e_{yy})\dot e_{xy}+\rm{2}\it \dot e_{yz}\dot e_{xz}]}.
\label{eq:sol-B}
\end{equation}
Here, $\alpha_1$ and $\alpha_2$ are functions of the physical properties of the fluid.
For Newtonian fluids, the stress is proportional to the strain rate.
Therefore, Eqs.~(\ref{eq:sol-A}) and (\ref{eq:sol-B}) can be expressed using stress as follows \citep{Nakamine2023}:
\begin{equation}
\begin{split}
\Delta_n\rm {cos}2&\phi=C_1({\sigma_{xx}}-{\sigma_{yy}})\\&+C_2[(\sigma_{xx}+\sigma_{yy})(\sigma_{xx}-\sigma_{yy})+\sigma_{zy}^2-\sigma_{xz}^2],
\end{split}
\label{eq:sol-A'}
\end{equation}
\begin{equation}
\Delta_n\rm{sin}2\phi=2\it{C_{\rm1}\sigma_{xy}+C_{\rm 2}[\rm{2}\it(\sigma_{xx}+\sigma_{yy})\sigma_{xy}+\rm{2}\it \sigma_{xz}\sigma_{yz}]}.
\label{eq:sol-B'}
\end{equation}
In these equations, $C_1=\alpha_1/\eta$ and $C_2 = \alpha_2/\eta^2$, in which $\eta$ is the shear viscosity of the fluid.
\cite{Aben1997} discussed the optical relationship based on Eqs.~(\ref{eq:sol-A'}) and (\ref{eq:sol-B'}) and assumed that the stress components along the camera's optical axis (hereafter simply the ``optical axis''), i.e., $\sigma_{xz}$, $\sigma_{zy}$, and $\sigma_{yz}$, were negligible.
In other words, they made the assumption that $C_2 = 0$, which leads to the proposal of:
\begin{equation}
    \Delta_n = C_1\sqrt{(\sigma_{xx}-\sigma_{yy})^2+4\sigma_{xy}^2}.
\label{stress-optic}
\end{equation}
From Eq.~(\ref{stress-optic}), it is clear that $\Delta_n=0$ in regions with $\sigma_{xy}=0$, such as the center of a channel flow.
However, non-zero $\Delta_n$ vales have been observed even when a quasi-two-dimensional channel was used \citep{Ober2011}.
This can be explained as being due to the shear caused by the upper and lower surfaces of the channel wall.
This discrepancy between theory and experiment is more significant in the case of three-dimensional flows, where the contribution of the shear components along the optical axis is larger \citep{Kim2017, Nakamine2023}.


Investigations of the response of flow birefringence to the shear rate (shear stress) have been conducted based on rheo-optical measurement techniques, which measure stress and flow birefringence simultaneously \citep{Ito2016, Lane2022}.
Measurement of flow birefringence is a potentially powerful tool for estimating the stress fields in fluids, and there have been some cases of successful reconstruction of shear-velocity fields in two-dimensional channel flows \citep{Kim2017}.
However, as Eq.~(\ref{stress-optic}) shows, the birefringence induced by the stress components along the optical axis has so far been neglected.
Therefore, there have been few cases of systematic measurements using an experimental setup in which a shear-velocity distribution exists along the optical axis, e.g., in a parallel-plate-type (PP-type) rheometer.
Such rheo-optical measurement setups are mainly applied to molten polymers \citep{Mykhaylyk2016}, although a complicated process is required to distinguish the effect of normal stress.
However, only qualitative measurements have been conducted for fluids with low viscoelasticity, e.g., dilute aqueous cellulose nanocrystal (CNC) suspensions \citep{Kadar2020}, and these are of great importance because they can be regarded as Newtonian fluids.

This study quantitatively investigated the flow birefringence induced by the shear components along the optical axis with respect to the shear rate.
For fluids with low viscoelasticity, the shear component along the optical axis can be directly visualized as birefringence.
Furthermore, herein, the trend of birefringence with respect to shear is discussed by comparing the present results---which involve birefringence induced by shear along the optical axis---with previous measurements of birefringence parallel to the shear.

\vspace{-8mm}
\section{Method}
\vspace{-5mm}
\subsection{Experimental setup}
\begin{figure}[tb]
\begin{center}
\includegraphics[width=0.47\textwidth]{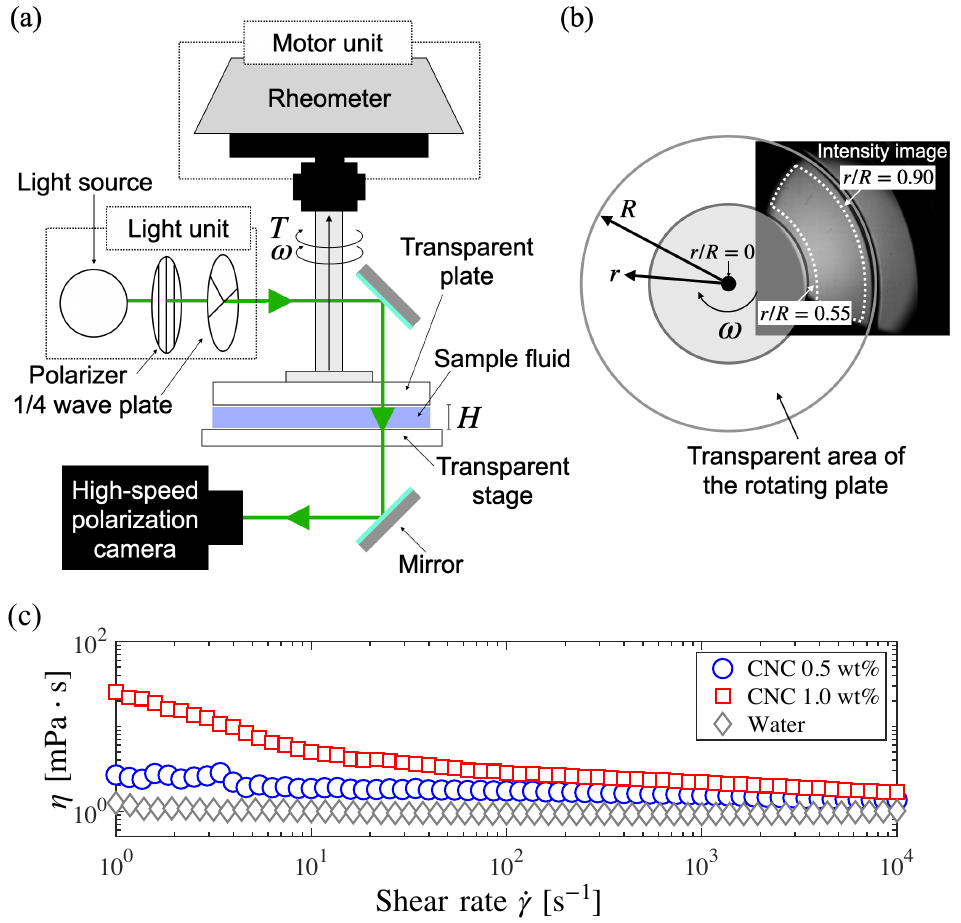}
\caption{(a)~Schematic of the experimental setup and (b)~an example image from the bottom view with an overlaid diagram showing the rotating plate, in which $R$ is the radius of the plate. (c)~Plots of steady shear viscosity $\eta$ for CNC suspensions at different concentrations.}
\label{fig:setup}
\end{center}
\end{figure}

\vspace{-4mm}
A PP-type rheometer (MCR~302, Anton Paar Co., Ltd.) was used for the polarization measurements in these experiments; a schematic of the experimental setup is shown in Fig.~\ref{fig:setup}a, and an example bottom-view intensity image is shown in Fig.~\ref{fig:setup}b.
A transparent plate (PP43/GL-HT, Anton Paar Co., Ltd.) and a stage (PTD200/GL, Anton Paar Co., Ltd.) were used to transmit the circularly polarized light emitted from the light source (SOLIS-525C, Thorlabs Co., Ltd., wavelength $\lambda = 525$~nm) through the flow.
The gap height between the plate and the stage was fixed at $H = (100 \pm 5)$~$\mu$m.
For polarization measurements, the transmitted light was captured from the bottom of the plate using a high-speed polarization camera (CRYSTA PI-1P, Photron Co., Ltd.) at 1000~frames per second.
The fluid temperature was maintained at 25$^\circ$C.
The range of the shear rate $\dot\gamma$ applied to the fluid was 0--10,000~s$^{-1}$.
For $r/R \geq 0.90$, a mottle-like birefringence distribution was observed even for optically isotropic liquids such as water.
This is induced by the refraction and scattering of light near the plate boundary.
In the present experiments and analysis, we focused on the region $0.55 \leq r/R\leq 0.90$, which is shown by the white dotted line in Fig.~\ref{fig:setup}b.
To evaluate the flow birefringence in a steady-state condition, temporal averaging was performed in the period 5.0--6.0~s after the plate started to rotate.
The retardation measured at $\dot\gamma = 0$~s$^{-1}$ was
uniformly offset from all other data.

\vspace{-8mm}
\subsection{CNC suspensions}
\label{CNC}
\vspace{-4mm}
Suspensions of CNC (Alberta Pacific Co., Ltd.) of two different concentrations were studied: 0.5 and 1.0~wt\%.
As shown in Fig.~\ref{fig:setup}c, the 0.5~wt\% CNC suspension behaved like a Newtonian fluid, whereas the 1.0~wt\% CNC suspension showed weak shear-thinning.
The normal stress was also measured using a rheometer, and the measured values were small enough to be regarded as measurement errors.
Note that the CNC suspensions of both concentrations had negligible elasticity.


\vspace{-8mm}
\subsection{High-speed polarization measurements}
\vspace{-4mm}
The high-speed polarization camera was used to detect the retardation $\Delta$, which is the integrated value of birefringence along the optical axis of the light transmitted through the apparatus. Using the phase-shifting method \citep{Onuma2014}, this was obtained from the radiance through linear polarizers oriented in four different directions (0$^\circ$, 45$^\circ$, 90$^\circ$, and 135$^\circ$) in an area of $2 \times 2$ pixels.
The light intensities detected at each of these pixels are defined as $I_1$, $I_2$, $I_3$, and $I_4$, respectively.
The retardation can then be expressed by \citep{Onuma2014}:
\begin{equation}
     \Delta = \int\Delta_n \mathrm{d}z=\frac{\lambda}{2\pi}{\rm sin^{-1}}\frac{2\sqrt{(I_3-I_1)^2+(I_2-I_4)^2}}{I_1+I_2+I_3+I_4},
\end{equation}
where $\lambda$ [m] is the wavelength of the light source.
In this study, we calculated the birefringence $\Delta_n$ by dividing the measured retardation by the gap height $H$.

\vspace{-8mm}
\section{Results and discussion}
\begin{figure}[tb!]
\begin{center}
\includegraphics[width=0.47\textwidth]{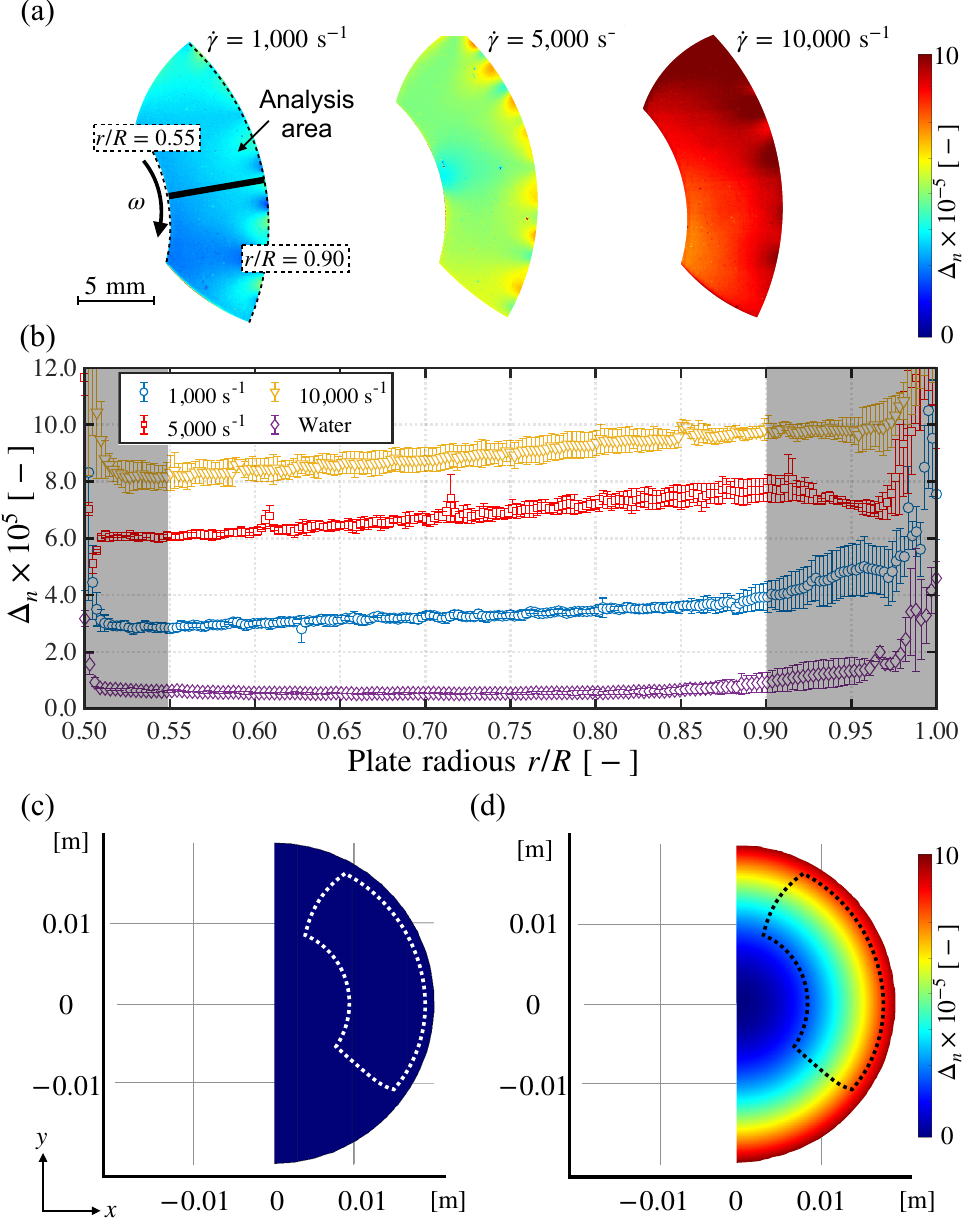}
\caption{(a)~Visualized birefringence fields under steady-state conditions; (b)~birefringence distribution of 1.0~wt\% CNC along the plate radius at each shear rate (error bars indicate standard deviations); analysis results of $\Delta_n$ when (c)~$C_2 = 0$~Pa$^{-2}$ and (d)~$C_2 = 2.0\times 10^{-7}$~Pa$^{-2}$. The areas enclosed by the dotted lines in panels (c) and (d) correspond to the areas shown in (a).}
\label{fig:All}
\end{center}
\end{figure}

\vspace{-4mm}
Visualized birefringence fields at different shear rates are shown in Fig.~\ref{fig:All}a.
As can be seen, the birefringence increased significantly as the shear rate increased.
Profiles taken across the section shown by the thick black line in the left-hand panel of Fig.~\ref{fig:All}a for different shear rates are plotted in Fig.~\ref{fig:All}b.
The shear rate increases outwardly from the center of the plate, which leads to an increase in the birefringence.

To discuss the mechanism of the birefringence induced by the shear stress, the experimental results (Figs.~\ref{fig:All}a and \ref{fig:All}b) were compared with the birefringence values calculated using Eq.~(\ref{stress-optic}).
When the optical axis is parallel to the $z$ axis, the stress components of the Couette flow between the rotating plate and the stage can be derived from the Navier--Stokes equations as
\begin{equation}
\sigma_{xx}=\sigma_{yy}=\sigma_{zz}=\sigma_{xy}=0.
\label{eq:sigma}
\end{equation}
As shown in Fig.~\ref{fig:All}c, the birefringence induced by the shear-stress loading, which was calculated using Eqs.~(\ref{stress-optic}) and (\ref{eq:sigma}), is $\Delta_n = 0$.
This indicates that the birefringence is induced by the stress components along the optical axis, which were ignored in Eq.~(\ref{stress-optic}).
In other words, this result suggests that the assumption that $C_2 = 0$ should not be applied to velocity fields with significant shear components along the optical axis, e.g., three-dimensional channel flows.
This means the birefringence measured in this study is a function of the stress components along the optical axis and $C_2$, which should be described as
\begin{equation}
\Delta_n = f(C_2,\ \sigma_{xz},\ \sigma_{zy},\ \sigma_{yz}).
\label{eq:C2}
\end{equation}

To further investigate the details of $C_2$, the birefringence was calculated using Eq.~(\ref{eq:C2}), and the results are shown in Fig.~\ref{fig:All}d.
Since values of $C_2$ for CNC suspensions have not been reported, the value was determined by fitting the absolute value of birefringence shown in Fig.~\ref{fig:All}b at 10,000~s$^{-1}$.
The fitted value was found to be $C_2 = 2.0\times10^{-7}$~Pa$^{-2}$.
Fitting was also conducted in different cases, and it was verified that the magnitude of $C_2$ was $O(10^{-7}$--$10^{-6})$.
Note that $C_1$ values for fluids have been reported to be $O(10^{-7}$--$10^{-5})$ in previous studies \citep{Ito2016,Nakamine2023}.
Although their units are different and therefore need to be discussed, $C_1$ and $C_2$ were found to be of similar magnitudes.
In addition, the distribution of birefringence in the radial direction of the plate is consistent with the results shown in Fig.~\ref{fig:All}b.

\begin{figure}[tb!]
\begin{center}
\includegraphics[width=0.50\textwidth]{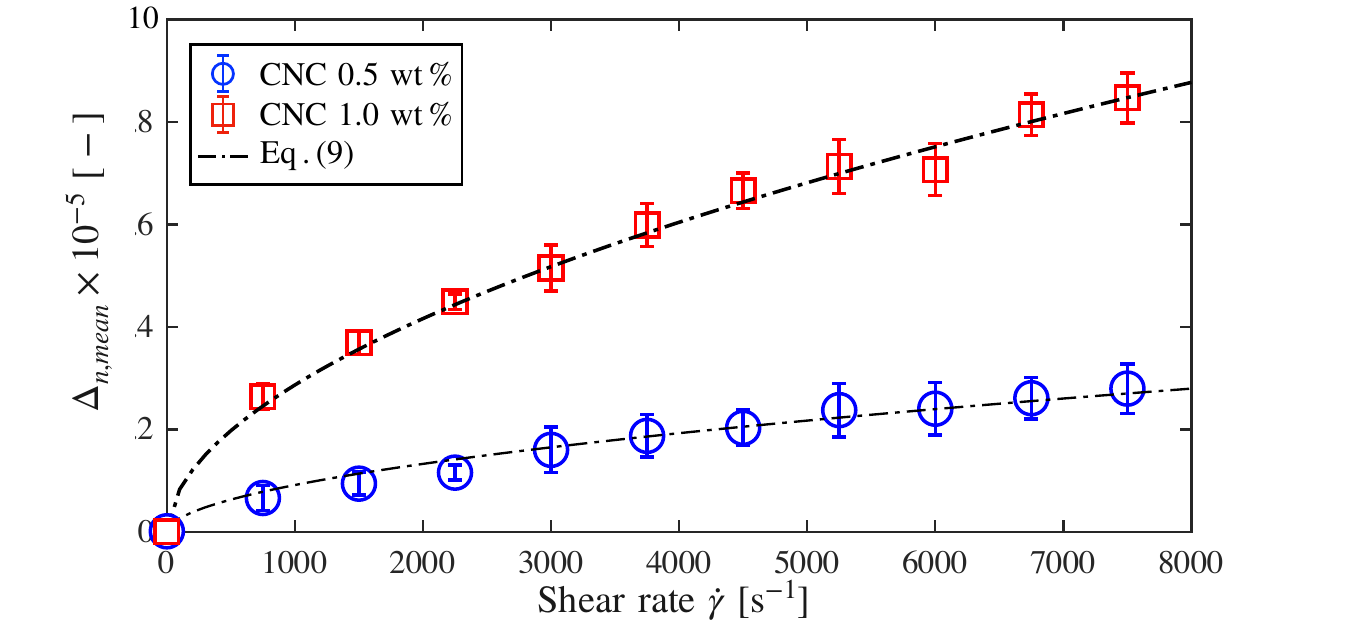}
\caption{Spatiotemporally averaged birefringence $\Delta_{n,{\it mean}}$ and corresponding fit using Eq.~(\ref{eq:fit}) with $A = 0.36\times 10^{-11}$~s, $k_1 = 0.538$, and $k_2 = 1.65$. The error bars indicate standard deviations.}
\label{fig:Fitting}
\end{center}
\end{figure}

Next, to validate the experimental birefringence results, the trend with respect to the shear rate was investigated.
In Fig.~\ref{fig:Fitting}, the vertical axis shows the spatiotemporally averaged birefringence $\Delta_{n,{\it mean}}$, while the horizontal axis shows the shear rate at $r/R = 0.75$.
When modeling the relationship between flow birefringence and shear rate, \cite{Lane2022} proposed that it can be described in the following nonlinear form:
\begin{equation}
    \Delta_n=(A\cdot \dot\gamma)^{k_1}\cdot c^{k_2},
    \label{eq:fit}
\end{equation}
where $c$ is the concentration of the suspension, and $A$, $k_1$, and $k_2$ are fitting parameters.
It should be emphasized that this model is based on the results of polarization measurements conducted from the direction parallel to the shear using a concentric-cylinder-type rheometer, which is different from that used in the present study.
The experimental results were fitted using Eq.~(\ref{eq:fit}), and the results are shown by the black dash-dotted lines in Fig.~\ref{fig:Fitting}.
The fitting parameters were $A = 0.36\times 10^{-11}$~s, $k_1 = 0.538$, and $k_2 = 1.65$.
Remarkably, the exponent $k_1$ (which characterizes the trend of the birefringence with the shear rate) obtained in this experiment has a similar value to that found in a previous study ($k_1 =0.537$, \cite{Lane2022}).
Furthermore, our results demonstrate the validity of the value of the exponent $k_1$ in a shear-rate range (0--7500~s$^{-1}$) that is much wider than that considered in the previous study (0--31.4~s$^{-1}$, \cite{Lane2022}).
The present experimental results and those of \cite{Lane2022} indicate that regardless of the direction of polarization measurement with respect to shear, there seems to be a common physical background that leads to the flow birefringence producing a power law for the shear rate.

\vspace{-10mm}
\section{Conclusion}
\vspace{-4mm}
In this letter, rheo-optical measurements were conducted on dilute CNC suspensions using a PP-type rheometer.
The novelty of this study lies in the fact that the birefringence was investigated from the direction perpendicular to the shear.
The shear rate was measured over a much wider range than in a previous study.
The measured birefringence was induced by the shear components along the optical axis, which was not considered in the stress--optic law, as $C_2 = 0$ in Eqs.~(\ref{eq:sol-A'}) and (\ref{eq:sol-B'}).
Our results indicate that birefringence induced by the shear components along the optical axis needs to be considered, especially for three-dimensional channel flows.
Additionally, they also suggest that the birefringence can be modeled by a power-law relationship with shear rate, similar to the results of a previous study in which polarization measurements were conducted from the direction parallel to the shear \citep{Lane2022}.
We are convinced that these findings are important for the development of a non-invasive fluid stress-field measurement method using flow birefringence.



\bibliographystyle{spbasic}
\bibliography{Refarence}

\end{document}